\documentstyle[12pt,epsf]{article}
\makeatletter

\renewcommand{\section}{\@startsection {section}{1}{\z@}
{-3.5ex plus -1ex minus -.2ex}{2.3ex plus .2ex}{\normalsize\bf}}

\renewcommand{\subsection}{\@startsection{subsection}{2}{\z@}
{-3.25ex plus -1ex minus -.2ex}{1.5ex plus .2ex}{\normalsize\it}}

\def\abstract{\if@twocolumn\section*{\abstractname}\else \small
\quotation\fi}\def\endabstract{\if@twocolumn\else\endquotation\fi}

\renewcommand{\@makefnmark}{\hbox{\mathsurround=0pt$^\dagger$}}

\renewcommand{\@makefntext}[1]{\parindent=1em\noindent
\hbox to 1.8em{\hss$^\dagger$}#1}

\def\thebibliography#1{\section*{\refname\@mkboth
 {\uppercase{\refname}}{\uppercase{\refname}}}\list
 {[\arabic{enumi}]}{\settowidth\labelwidth{[#1]}\leftmargin\labelwidth
 \advance\leftmargin\labelsep\parsep=0em\itemsep=0em
 \usecounter{enumi}} \def\newblock{\hskip .11em plus .33em minus .07em}
 \sloppy\clubpenalty4000\widowpenalty4000
 \sfcode`\.=1000\relax}\makeatother
\begin{document}
\renewcommand{\H}{{\cal H}}
\newcommand{\N}{{\cal N}}
\newcommand{\Z}{{\cal Z}}
\newcommand{\D}{{\cal D}}

\begin{center}
{\normalsize\bf
A GENERALIZATION OF THE SPHERICAL MODEL:
Reentrant phase transition in the spin one ferromagnet.
}
\\
\bigskip
S.E. Savel'ev$^{a,}$
\footnote{Corresponding author. Fax: +7 (095) 3619226;
e-mail: sesavelev@glasnet.ru}
and Guillermo Ram\'\i rez-Santiago$^b$
\medskip \\
{\small\it
$^a$
All-Russian Electrical Engineering Institute, Moscow 111250,
Russia
\\$^b$
Instituto de Fisica, UNAM, PO Box 20-364 M\'exico, 0100,
D.F. MEXICO
}
\end{center}

\begin{abstract}
By mapping the hamiltonian of the spin one ferromagnet onto that of the
classical spherical model we investigate the possible phase transitions and
the phase diagram of the spin one ferromagnet. Similarly to what happens in
the spherical model we find no phase transition in one and two dimensions.
Nonetheless, for three dimensions we obtain a phase diagram in which the most
important and unexpected feature is the existence of a
paramagnetic-ferromagnetic-paramagnetic transition at low temperatures.
\end{abstract}

\section{Introduction}
The spin one ferromagnet has three states $S_{z}=0,\pm1$ and can
be used as a model of several interesting physical systems.
For example, it can be
applied to the qualitative study of the $\lambda$ transition and phase
separation of $^{3}$He-$^4$He\cite{beg} mixtures.
In addition, it can model an Ising ferromagnet $S_{z}=\pm {{1}\over{2}}$ with
nonmagnetic impurities $S_{z}=0$ as well as a magnetic system with competing
interactions\cite{jensen87}.
It can also model the behavior of a microemulsion where the interplay of
the components, water, oil and amphiphile, is mimicked by the
states $S_{z}=0,\pm1$\cite{schick87}.
Thus, studying the phase transitions and the phase diagram of the spin
one ferromagnet has physical relevance.
In this paper we analyze the properties  of the spin one ferromagnet by
introducing a transformation that maps the model hamiltonian onto that of
the classical spherical model (CSM). Then  we  study the properties of the
saddle points of the integrand of the partition function and look for the
existence of phase transitions. The main result of this paper is the
existence of a reentrant
pa\-ra\-mag\-ne\-tic-fe\-rro\-mag\-ne\-tic-pa\-ra\-mag\-ne\-tic transition.
Consider a $d$-dimensional cubic lattice with three states
$N_{\vec R}=0,\pm 1$ spin located at each of its nodes.
Due to symmetry reasons we may assume that the states $N_{\vec R}=\pm 1$
are equivalent energetically, while the state $N_{\vec R}=0$ may have
different energy. The physical motivation for these assumptions comes from
the systems mentioned above.
Assuming that the spins interact through the pair potential
$V(\vec R-\vec R')$ and with an  external magnetic field $H$, then
the general Hamiltonian that defines the  model  is written as
\begin{equation}
\H=E\sum_{\vec R}N_{\vec R}^2+\sum_{\vec R, \vec R'}
V\left({\vec R}-{\vec R'}\right)N_{\vec R}N_{\vec R'}
+H\sum_{\vec R}N_{\vec R},
\label{ham0}
\end{equation}
where the second sum is over all pairs of spins.
The first sum in $\H$ has been introduced in order to have the
possibility of quadrupolar ordering \cite{beg} in addition to the magnetic
ordering associated with the second sum. The competition of these two
types of ordering is measured by the parameter $E$, the energy difference
between the states with $N_{\vec R}=\pm 1$ and $N_{\vec R}=0$.
With the aim at mapping the model hamiltonian onto that of the spherical
model we introduce the Ising-like variables $a_{\vec R}=\pm 1$ and
$b_{\vec R}= \pm 1$ that are related to the original spin variables by,
$N_{\vec R}=\frac{1}{2}\left(a_{\vec R}+b_{\vec R}\right)$.
Under this transformation the state $N_{\vec R}=0$ has been double weighted
since it can be obtained from the states $a_{\vec R}=-b_{\vec R}= 1$ and
$a_{\vec R}=-b_{\vec R}=-1$.
To compensate this difference we introduce the additional entropy term
$ {\cal S}=-k_{B}T\ln 2\sum_{\vec R}N^{2}_{\vec R}$, with
$k_{B}=1$ the Boltzmann constant.  In terms of the Ising-like variables
$\H$ becomes:
\begin{eqnarray}
\H &=& \frac{1}{4}(E-T\ln 2) \sum_{\vec R}\left(a_{\vec R}+b_{\vec R}\right)^2
+\frac{1}{4}\sum_{\vec R, \vec R'}V\left({\vec R}-{\vec R'}\right)
\left(a_{\vec R}+b_{\vec R}\right)
\left(a_{\vec R'}+b_{\vec R'}\right)\nonumber\\
&+& \frac{H}{2}\sum_{\vec R}\left(a_{\vec R}+b_{\vec R}\right).
\label{ham1}
\end{eqnarray}
To study analytically the properties of the model we replace the discrete Ising-variables
by continuous ones $-\infty < a_{\vec R}, b_{\vec R} < +\infty$, subject
to the spherical conditions:
$\sum_{\vec R}a_{\vec R}^2=\N$, and $\sum_{\vec R}b_{\vec R}^2=\N$,
with $\N$ the total number of sites (spins) in the system. Hence,
the hamiltonian (\ref{ham1}) together with the last two restrictions
can be considered as the spherical version of the original hamiltonian.

\section{Results}
Proceeding  as in the CSM \cite{berlin-kac} the partition function
$\Z$, after some straightforward calculations, can be written as
\begin{equation}
\Z=\pi^\N\int\frac{ds_1 ds_2}{(2\pi i)^2}\exp{\N\gamma(s_1, s_2)}
\label{int-z}
\end{equation}
where
\begin{equation}
\gamma\left(s_1, s_2\right)=s_1+s_2-\frac{1}{2\N}\sum_{\vec k}
\ln\left( s_1s_2+\left(s_1+s_2\right)\frac{\alpha_{\vec k}}
{T}\right)
+\frac{H^2}{16 T^2}\frac{s_1+s_2}{\frac{\alpha(0)}{T}
\left(s_1+s_2\right)+s_1s_2}
\end{equation}
and
$\alpha\left(\vec k \right)=\frac{1}{4}\left[ E-T\ln 2+
\sum_{\vec R}V\left(\vec R\right)\exp(-i\vec k\vec R)\right]$.
The integral in Eqn. (\ref{int-z}) can be approximated by evaluating
the integrand at its saddle points: $\frac{\partial\gamma}{\partial s_i}=0$
with $i,j=1,2$ and $i\neq j$. The solution $s_i=s_j=s$ is  a root of
the equation,
\begin{equation}
\psi (s)=\frac{1}{\N}\sum_{\vec k}\frac{1}{s+2\frac{\alpha(\vec k)}
{T}}+\frac{H^2}{4T^2}\frac{1}{\left(s+2\frac{\alpha(0)}{T}\right)^2}+
\frac{1}{s}=4
\label{ss}
\end{equation}
If $H\ne 0$ there is a saddle point at any temperature only if
$\max_{\vec k}\alpha(\vec k)=\alpha(0)$ which cannot be accomplished.
On the other hand, for $H=0$ a  saddle point may disappear if in the limit
$s\rightarrow -2\alpha(0)/T$ the sum converges to a finite value.

Let us consider a system where $V(\vec R-\vec R')=-J$ and $\vec R,\vec R'$
are nearest neighbors.
Bearing in mind that for the CSM the absence (existence) of
a saddle point is related to a ferromagnetic (paramagnetic) phase
and assuming that $H=0$, Eqn.~(\ref{ss}) can be rewritten as
\begin{equation}
4=\psi\left(s\right)=\frac{1}{s}+\frac{T}{J}\frac{1}{(2\pi)^d}
\int_{B.Z.} d\vec q\frac{1}{\frac{T}{J}\left(s+\frac{E}{2T}-
\frac{1}{2}\ln 2\right)-\left(\sum_{j=1}^{d} \cos q_j\right)}
\label{sss}
\end{equation}
where we have replaced the sum by an integral over the Brillouin zone.
The function $\psi(s)$ decreases monotonically and is analytic for
$s_0(J/T) < s \leq +\infty$ with
$s_0=\max\left(0, \frac{1}{2}\ln 2 -(E-2dJ)/2T\right)$.
Thus, a solution (a saddle point) of Eq.~(\ref{sss}) exists if $\psi(s_0)>4$
(paramagnetic phase),
while there is no saddle point if $\psi(s_0)<4$ (ferromagnetic phase).
For $d=1$ it is easy to show that when $s\rightarrow s_0$,
$\psi (s)$ diverges yielding no phase transition. For $d=2$ the evaluation
of such integral is similar to that of the CSM and one can arrive to the
conclusion that there is no phase transition \cite{berlin-kac}.

In $d=3$  for some values of the parameters $E$ and $J$, $\psi (s_0) < 4$
and there is no saddle point in the integrand of $\Z$, leading to a phase
transition. A careful analysis of $\psi \big[s_0(T),T\big]$ indicates
that there are two ranges of parameters to be considered:
(1) $E-6J<0$ in which case $s_0(T)$ goes from $+\infty$ at $T=0$ up to
$\frac{1}{2}\ln 2$ at $T=\infty$,
and (2) $E-6J>0$ for which  $s_{0}(T)$ goes from zero at $T=0$ up to
$\frac{1}{2}\ln 2$ at $T=\infty$. Here following ref. \cite{berlin-kac}
we  use the result
$({{1}\over{2\pi}})^{3}\int d^{3}q\Big(3-\sum_{j=1}^{3}
\cos q_j\Big)^{-1} \approx 0.50541 = I_{\rm cr}$.
Therefore $\psi(s_0)$ can be rewritten as
\begin{equation}
\psi(s_0(T), T)=\frac{T}{J}I_{\rm cr}+\frac{2T}{T\ln 2-E+6J}=4.
\label{psi0}
\end{equation}
At very low temperature and for $E-6J<0$, the parameter 
$s_{0}$ is quite large  and
$\psi(s_{0})$ goes down to zero in which case the system settles in the
ferromagnetic state  ($\psi(s_{0})<4$).
Nonetheless, at high temperatures $s_{0}$ approaches to
${ {1}\over {2}}\ln  2$  ($\psi(s_0(T),T)>4$) yielding the paramagnetic
phase.
In Fig. 1(a) we show the behavior of $\psi(s_0(T), T)$ 
as function of temperature
for several values of $E/J$. There we see that for
$E/J<6.0$ the function $\psi(s_0(T), T)$ increases monotonically as $T$
grows starting from $T=0$ up to the high temperature region.  On the other
hand, for $E/J > 6.0$ and low temperatures it decreases until a minimum is
reached at $T^{*}=(1/\ln 2) \Big( E-6J+\sqrt{2J(E-6J)/I_{\rm cr}}\>\> \Big)$,
and then increases and goes asymptotically to the line $\psi(E/J=6,T)$ at
higher temperatures. The critical line
defined by $\psi(s_0(T_c), T_c)=4$, horizontal dotted line in Fig. 1(a),
intersects  the curves $\psi(s_0(T),E/J)$ in only one point if $(E/J)<6$,
single critical temperature, while for  $6<(E/J)\leq 6+x_{c}$
it intersects $\psi(s_0(T),E/J)$ in two points giving rise to two 
critical temperatures.
Here $x_{c}=(1/I_{cr})(2\sqrt{\ln 2} - \sqrt{2})^{2}\approx 0.124 \>$.
These curves will be denoted as {\it reentrance lines} for values of $E/J$ in
$[6,6+x_{c}]$.
In this range of parameters it is also found that the paramagnetic phase 
($\psi(s_0(T),T)>4$) sets in at
low and high temperatures while the ferromagnetic one 
($\psi(s_0(T),T)<4$) only exists at the
intermediate regime. Above the upper bound $(E/J)_{c}=6+x_{c}$,
there will be no phase transition at all. The critical temperatures
obtained as the roots of the critical line are:
\begin{equation}
T^{\pm}_{c}=\frac{(E-6J)I_{\rm cr}+4J\ln 2-2J \pm \D}
{2I_{\rm cr} \ln 2}
\label{Tc}
\end{equation}
with
$\D=\Big[ \big( (E-6J)I_{\rm cr}+4J\ln 2-2J\big)^2
-16(E-6J)I_{\rm cr} J\ln 2 \Big]^{1/2}$  These two branches are plotted
as function of $E/J$ in Fig. 1(b). The upper branch $T^{+}_{c}(E/J)$ goes
from $2.5J \leq T_{c} < 5.5J$ while the lower branch $T^{-}_{c}(E/J)$
extends  from $T_{c}=0$ up to $T_{c}=2.5J$. In the inset of Fig. 1(b) we show
the {\it reentrance line},
$T_{c}(E/J)$ for $6 \leq {{E}\over{J}} \leq 6+x_{c}$.

At this point it is important to note that the results presented here are
stable for small changes of the entropy term
$ {\cal S}=-k_{B}T\ln 2\sum_{\vec R}N^{2}_{\vec R}$
introduced as an extra term in the original $\H$.

\section{Conclusions}
In this paper we have shown analytically that the spherical version of the
spin one ferromagnet with short range interactions and zero external
magnetic field  shows a reentrant phase transition to the paramagnetic
state at low temperatures in $d=3$ for $6<(E/J) \leq 6.124$.  However, in
lower dimensions we find no phase transition as it occurs in the spherical
model. We also obtained the phase diagram $T_{c}$ versus $E/J$ where the
different phases are indicated.  This mapping of the system onto its
spherical version has proven to be very useful since it may help to shed
light upon the physics of several interesting systems \cite{ran-inter}.
For instance, we are currently applying this approach to the study of
phase transitions in Josephson junctions arrays with disorder
\cite{aps99}.  An extended version of the present  paper that includes
more details as well as additional results will be published elsewhere
\cite{pre99}.

This work has been supported by CONACYT-MEXICO under contract 25298-E.

\begin{figure}
\epsfxsize=14cm
\epsfysize=10cm
\centerline{\epsfbox{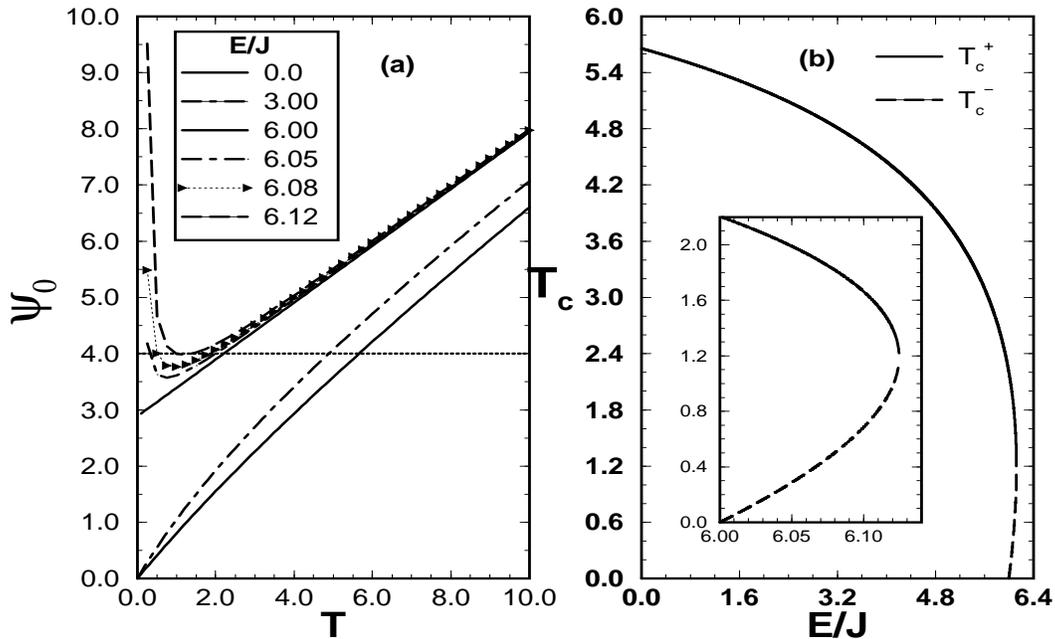}}
\caption[]{ See text for an explanation of these figures.}
\end{figure}

\end{document}